\documentclass[aps,prl,twocolumn,floatfix,showpacs,longbibliography,superscriptaddress]{revtex4-2}
\usepackage{amsmath}
\usepackage{amssymb}
\usepackage{times}
\usepackage[pdftex]{graphicx}
\usepackage{color}
\usepackage[colorlinks=true, citecolor=blue, urlcolor=blue, linkcolor=blue]{hyperref}
\usepackage{tikz-cd}
\usepackage{babel}
\newcommand{\bu}{\mathbf{u}}
\newcommand{\br}{\mathbf{r}}
\newcommand{\bn}{\mathbf{n}}
\newcommand{\bk}{\mathbf{k}}

\begin{document}

\preprint{APS/123-QED}

\title{Construction of Hopfion Crystals}

\author{Wen-Tao Hou}
\email{houwentao@tiangong.edu.cn}
\author{Zhuoxian Xiang}
\affiliation{School of Physical Science and Technology, Tiangong University, Tianjin 300387, China}

\author{Yizhou Liu}
\affiliation{Anhui Province Key Laboratory of Low-Energy Quantum Materials and Devices, High Magnetic Field Laboratory, HFIPS, Chinese Academy of Sciences, Hefei 230031, China}

\author{Jiadong Zang}
\email{jiadong.zang@unh.edu}
\affiliation{Department of Physics and Astronomy, University of New Hampshire, NH 03824, USA}

\date{\today}

\begin{abstract}
  Hopfions, three-dimensional topological solitons characterized by nontrivial Hopf indices, represent a fundamental class of field configurations that emerge across diverse areas of physics. Despite extensive studies of isolated hopfions, a framework for constructing spatially ordered arrays of hopfions, i.e., hopfion crystals, has been lacking. Here, we present a systematic approach for generating hopfion crystals with cubic symmetry by combining the Hopf map with rational mapping techniques. By superposing helical waves in $\mathbb{R}^4$, we construct hopfion crystals with tunable Hopf indices and controllable topology. We demonstrate simple cubic, face-centered cubic, and body-centered cubic hopfion crystals, and extend our framework to create crystals of more complex topological structures, including axially symmetric tori, torus links, and torus knots with higher Hopf indices. Our results provide a foundation for searching hopfions in real materials and studying their collective phenomena.
\end{abstract}



\maketitle

In 1975, Faddeev introduced hopfions as a class of truly three-dimensional (3D) topological solitons characterized by integer Hopf indices\cite{hopf_uber_1931,Faddeev1975Soliton,Faddeev1976soliton}. Since then, hopfions have sparked intense research interest and been studied in a variety of physical systems across scales, including optical fields\cite{Optical2022Shen1,Optica2023,ShenZhu2023,Optical2024ACS,Shen2024}, Bose-Einstein condensates\cite{BEC2015,BEC12015,li_three_dimensional_2016,BEC2021}, liquid crystals\cite{LiquidCrystal2015Hopfion,CrystalLiquid2022}, magnetic materials\cite{Frustrated2017Sutchliffe,ChiralNanodish2018,Fisher2021Hopfion,YuLiu2023FeGe,Zheng2023Nature}, ferroelectric materials~\cite{Hopf2020Ferro}, etc.
Hopfions, as 3D topological solitons, exhibit rich topological structures such as rings, links, and knots, which give rise to novel physical phenomena.
For example, in magnetic systems, theoretical studies have revealed hopfions' unique nonlinear transport effects and 3D dynamic characteristics\cite{Dynamic2019,Dynamics2020We,Dynamic2021,pershoguba_electronic_2021,Yizhou2022Hopfion,hou_microscopic_2024}, highlighting their potential applications in 3D spintronics\cite{3D_roadmap}. Recent advances in fabrication and three-dimensional imaging techniques from nano- to micro-scales have made it possible to access the details of hopfions in laboratories\cite{three_dimensional_2017,donnelly_imaging_2022,CrystalLiquid2022,3D_roadmap}.

Despite growing interest, searching for hopfions in real physical systems and investigating their collective behaviors remain challenging due to their sophisticated configurations and complex models. Obtaining analytical expressions for hopfion configurations is essential for studying their stability, dynamics, and physical properties.
While analytical ansatzes for single hopfions with various topologies have been developed based on methods like rational maps\cite{faddeev_stable_1997,Sutcliffe1998Map,Ferromagnet2023}, the theoretical framework for constructing spatially ordered hopfion crystals, critical for unlocking real material models that host hopfions as ground states and studying emergent phenomena in hopfion ensembles, has been lacking.

\begin{figure}
  \centering
  \[
    \begin{tikzcd}
      & \mathbb{R}^4  \arrow[d] & \\
      \mathbb{R}^3 \arrow[ur] \arrow[r, "sp"] & \mathbb{S}^3 \arrow[r, "rp"] & \mathbb{C} \arrow[r,"sp"] & \mathbb{S}^2
    \end{tikzcd}
  \]
  \caption{Two mapping pathways from $\mathbb{R}^3$ to $\mathbb{S}^2$. The lower one is the standard construction for a single hopfion, and the upper one is the path for hopfion crystals. $sp$ stands for sterographic projection, and $rp$ stands for rational map.}
  \label{fig:map}
\end{figure}
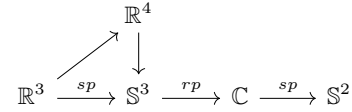

In this Letter, we present a method for constructing hopfion crystals, which offers a robust framework for generating topological solitons in three dimensions. We start by first reviewing the construction process of a single hopfion. A hopfion realizes the map from $\mathbb{R}^3$ to $\mathbb{S}^2$. As long as the spins at infinity are polarized to the same direction as required by the finite energy, the base manifold $\mathbb{R}^3$ is compactified to $\mathbb{S}^3$. The isomorphism between $\mathbf{r}\in\mathbb{R}^3$ and $\mathbf{u}\in\mathbb{S}^3$ is enabled by the stereographic projection
\begin{align}
  u_i & = \frac{r_i}{r}\sin f, \nonumber \\
  u_4 & = \cos f
  \label{eq:u_vector}
\end{align}
where $i=\{1,2,3\}$ are the cartesian coordinates. A common choice of the angle $f$ is $f=\arccos\frac{1-r^2}{1+r^2}$, but it can be any smooth function interpolating between $0$ and $\pi$ from origin to infinity. The Hopf map $\mathbb{S}^3\rightarrow\mathbb{S}^2$ can be constructed by the rational map\cite{Sutcliffe1998Map,kedia_weaving_2016}. Introducing two complex coordinates on $\mathbb{S}^3$
\begin{equation}
  Z_1 = u_1 + iu_2, \qquad Z_0 = u_3 + iu_4.
\end{equation}
One can construct a variable $\xi\in\mathbb{C}$ in the Riemann complex plane by
\begin{equation}
  \xi=\frac{P(Z_1,Z_0,\bar{Z}_1,\bar{Z}_0)}{Q(Z_1,Z_0,\bar{Z}_1,\bar{Z}_0)}
  \label{eq:rational_map}
\end{equation}
where $P$ and $Q$ are polynomials of $Z_1,Z_0,\bar{Z}_1$ and $\bar{Z}_0$ (the complex conjugates of $Z_1$ and $Z_0$). Depending on these two polynomials, different types of hopfions can be constructed. The final step is the stereographic projection $\mathbb{C}\leftrightarrow\mathbb{S}^2$:
\begin{equation}
  n_1=\frac{2\Re(\xi)}{1+\xi\bar{\xi}}, \quad n_2=\frac{2\Im(\xi)}{1+\xi\bar{\xi}}, \quad n_3=\frac{1-\xi\bar{\xi}}{1+\xi\bar{\xi}}
\end{equation}
or equivalently
\begin{equation}
  \xi = \frac{n_1+in_2}{1+n_3}.
\end{equation}
The isomorphism $\mathbb{C}\leftrightarrow\mathbb{S}^2$ can be also constructed in terms of a spinor
\begin{equation}
  \chi = \mathcal{N}\begin{pmatrix}
    Q \\P
  \end{pmatrix}, \qquad \mathbf{n}=\langle\chi|\mathbb{\sigma}|\chi\rangle.
  \label{eq:spinor_construct}
\end{equation}
Here $\mathcal{N}$ is the normalization factor. The U(1) gauge degree of freedom in $\chi$ is equivalent to projective degree of freedom in Eq.~\ref{eq:rational_map}. The entire mapping pathway for a single hopfion is shown in FIG.~\ref{fig:map}.

A simple choice for $\xi$ is $\xi=Z_1/Z_0$, corresponding to the spinor $|\chi\rangle=(Z_1,Z_0)^T$. Equi-spin contour, i.e, the preimage of map $\mathbb{R}^3\rightarrow\mathbb{S}^2$, of $\mathbf{n}=(0,0,-1)$ locates at $x=y=0$ where $\xi=\infty$. Meanwhile, on the preimage of $\mathbf{n}=(0,0,1)$, $\xi=0$ so that $u_3=u_4=0$, $z=\cos f=0$. If the conventional choice $\cos f=(1-r^2)/(1+r^2)$ is used, that corresponds to a circle with $r=1$ on the $x-y$ plane.

The topological index of the hopfion is the linking number between two arbitrary preimages. On the preimage $\xi$ is kept constant, so that its tangential direction defines an emergent magnetic field
\begin{equation}
  \mathbf{b} = \frac{1}{2\pi i}\frac{\mathbb{\nabla}\xi\times\mathbb{\nabla}\bar{\xi}}{(1+\xi\bar{\xi})^2}.
\end{equation}
It can be easily proved that this field is equivalent to the real space Berry curvature of the spin texture
\begin{equation}
  \mathbf{b}=\frac{1}{4\pi}\mathbf{n}\cdot(\mathbb{\nabla}\mathbf{n}\times\mathbb{\nabla}\mathbf{n}).
  \label{eq:topological_charge}
\end{equation}
The Hopf index is thus calculated as
\begin{equation}
  Q_H=\int d^3r\ \mathbf{a}\cdot\mathbf{b}
  \label{eq:hopf_index}
\end{equation}
given $\nabla\times\mathbf{a}=\mathbf{b}$ an arbitrary vector potential. A straightforward choice is $\mathbf{a}=-i\langle\chi|\nabla|\chi\rangle$. The example above has the Hopf index $Q_H=1$.

\begin{figure}
  \includegraphics[width = \columnwidth]{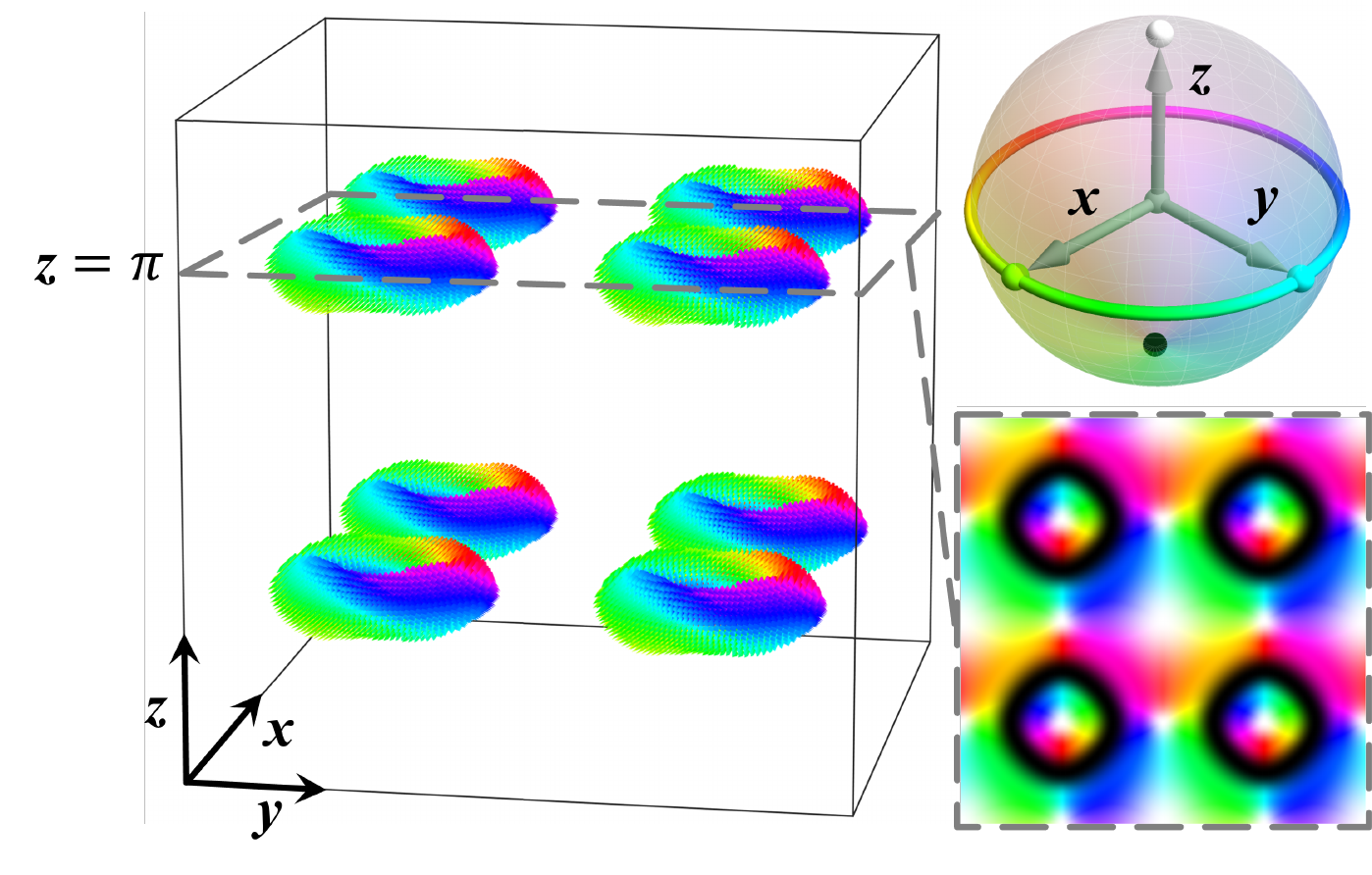}
  \caption{\label{fig:sc}$Q_{H}=1$ simple cubic hopfion crystal. The left figure is the simple cubic $Q_{H}=1$ hopfion crystal with $c=2$ and $x,y,z$ are from $-2\pi$ to $2\pi$. The right top figure indicates the the relationships between colors and  the values of $\arctan(n_{y}/n_{x})$ and $n_z$. The right bottom figure is the spin texture of $z=\pi$ plane.}
\end{figure}

Generalization of the aforementioned single hopfion to a hopfion crystal is inspired by the construction of the skyrmion crystal. A closely packed skyrmion crystal is a superposition of three spin helices: $\bn(\br)=\bn_0+\sum_{i=1}^{3}\bn(\bk_i)\exp(i\bk_i\cdot\br)+\text{c.c.}$\cite{muhlbauer_skyrmion_2009}. Here, three wave vectors $\bk_i$ having the same pitch length are co-planar and equal lateral. $\bn(\bk)=\frac{1}{2}(\hat{z}-i\hat{\bk})$ gives a N\'eel skyrmion crystal. A Bloch skyrmion crystal takes another choice of $\bn$ but it relates to the N\'eel skyrmion by $\pi/2$ rotation about $\hat{z}$ of all spins, so the topology is the same. $\bn_0$ is a constant vector associated with $\bk=0$. Its presence is essential to avoid singularity $\bn=0$. This construction is able to build a crystal with nonzero topological charge $\int d^2r\ b_z$ because the Fourier transformation of the emergent magnetic field in Eq.~\ref{eq:topological_charge} involves three independent $\bn(\bk)$ associated with three different $\bk$. The topological index of hopfion in Eq.~\ref{eq:hopf_index}, however, involves six $\bn(\bk)$. A direct search of valid $\bn(\bk)$ likes finding a needle in a haystack.

The key idea behind the skyrmion crystal is to embed a spin vector—normally defined on $\mathbb{S}^2$—into $\mathbb{R}^3$ instead. It can recover $\mathbb{S}^2$ space by renormalization. Therefore, an analogous way of constructing the hopfion crystal is to embed the vector $\bu$ in Eq.~\ref{eq:u_vector} into $\mathbb{R}^4$ space instead. The map from $\mathbb{R}^3$ to $\mathbb{S}^2$ takes the upper path in FIG.~\ref{fig:map}. Imitate the expression for the N\'eel type skyrmion crystal, we have
\begin{equation}
  \bu(\br) = \bu_0 + \sum_{i=1}^3 \bu(\bk_i)\exp[i(\bk_i\cdot\br+\phi_i)]+\text{c.c.}
\end{equation}
where $\phi_i$ are phase factors of each helix. Here three independent $\bk_i$ are taken. This is the minimal number of linearly independent vectors in three dimensions. In the skyrmion crystal in two dimensions, in principle, superposition of two wave vectors are enough to enable topological textures. However, the resulting state is a lattice of meron and anti-meron with zero net topological charge\cite{yu_transformation_2018,hayami_phase_2021}. In our current case, it will be shown shortly that three wave vectors are adequate.

\begin{figure*}
  \includegraphics[width = \textwidth]{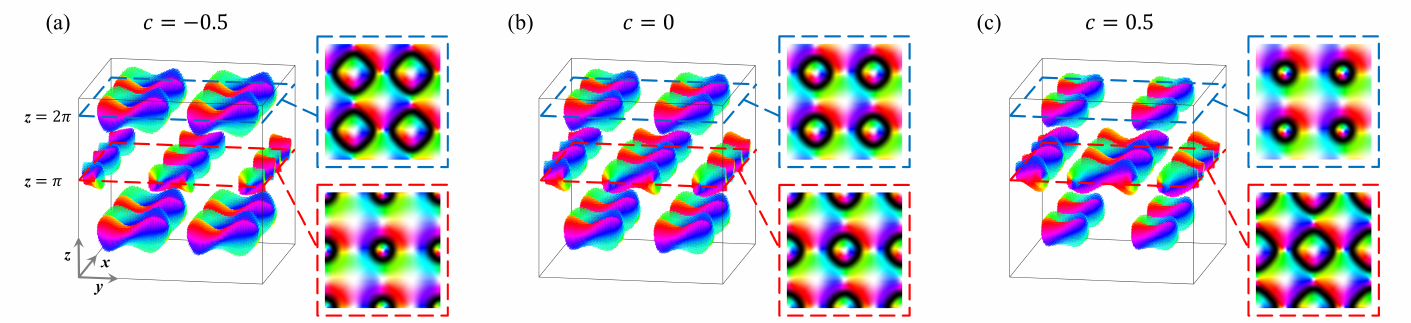}
  \caption{\label{fig:diff-c}Hopfion crystal structures with differnet $c$ values. (a) $c=-0.5$ hopfion crystal structures with the spin textures at $z=\pi$ and $z=2\pi$ planes. (b) $c=0$ hopfion crystal structures with the spin textures at $z=\pi$ and $z=2\pi$ planes. (c) $c=-0.5$ hopfion crystal structures with the spin textures at $z=\pi$ and $z=2\pi$ planes.}
\end{figure*}

We start with $\bk_i=\hat{r}_i$ and $\bu(\bk_i)=\frac{1}{2}(\hat{l}-i\hat{\bk}_i)$ in analogy to the skyrmion crystal. Here $\hat{l}$ is the unit vector in the fourth dimension. As a result, each helix takes the following forms
\begin{align}
  \bu_{1} & =(\sin({\bf k}_{1}\cdot {\bf{r}}+\phi_{1}),\ 0,\ 0,\ \cos({\bf k}_{1}\cdot {\bf r}+\phi_{1})),\nonumber \\
  \bu_{2} & =(0,\ \sin({\bf k}_{2}\cdot {\bf r}+\phi_{2}),\ 0,\ \cos({\bf k}_{2}\cdot {\bf r}+\phi_{2})),\nonumber  \\
  \bu_{3} & =(0,\ 0,\ \sin({\bf k}_{3}\cdot {\bf r}+\phi_{3}),\ \cos({\bf k}_{3}\cdot {\bf r}+\phi_{3}))
  \label{eq:helix}
\end{align}
and $\bu_0=(0,0,0,c)$ is introduced to avoid singularity at which $\bu=\bu_0+\bu_1+\bu_2+\bu_3=\mathbf{0}$. If a singularity were to exist, all sines in Eq.~\ref{eq:helix} are zero, and all cosines are $\pm1$. Therefore, $c$ cannot take the values of $\pm1$ and $\pm3$. Every time these values are crossed, the Hopf index changes.

With these preparations, a hopfion crystal can be constructed using the same rational map. We again choose the simple case with Riemann coordinate $\xi=Z_1/Z_0$. The corresponding spinor state is then given by
\begin{align}
  |\chi\rangle=\mathcal{N}
  \begin{pmatrix}
    \sin Q_{1}+i\sin Q_{2} \\
    \sin Q_{3}+i(\cos Q_{1}+\cos Q_{2}+\cos Q_{3}+c)
  \end{pmatrix}
  \label{eq:spinor}
\end{align}
where $Q_{i}={\bf k}_{i}\cdot{\bf r}+\phi_{i}$ and $\mathcal{N}=[(\cos Q_{1}+\cos Q_{2}+\cos Q_{3}+c)^{2}+\sin^{2}Q_{1}+\sin^{2}Q_{2}+\sin^{2}Q_{3}]^{-\frac{1}{2}}$ is the renormalization factor. Without loss of generality, $k$, the amplitude of each wave vector, is chosen to be 1.

\begin{figure}
  \includegraphics[width = \columnwidth]{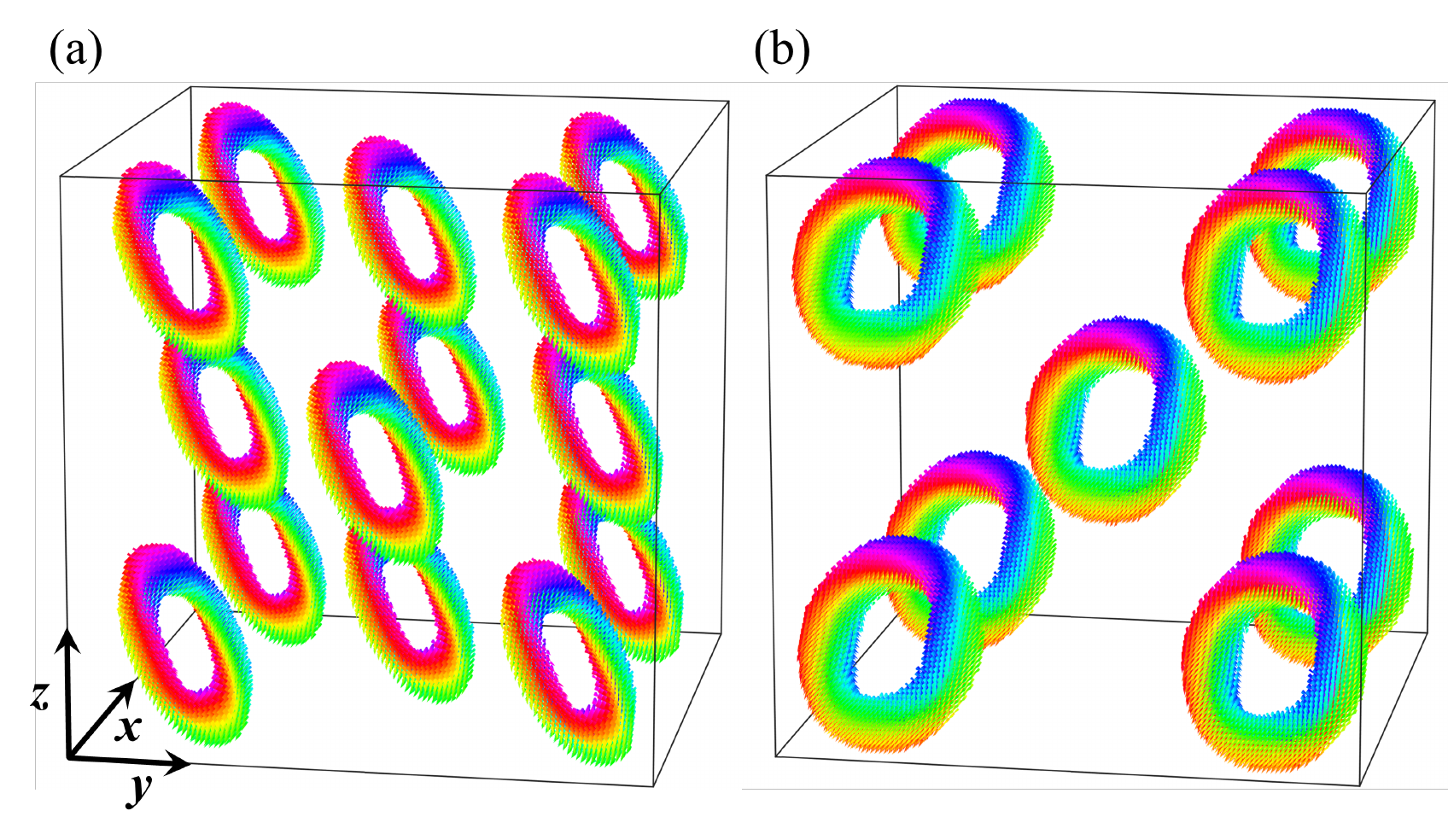}
  \caption{\label{fig:fcc+bcc}(a) $Q_{H}=1$ face-centered cubic hopfion crystal. (b) $Q_{H}=1$ body-centered cubic hopfion crystal. We make $c=2$ for both of them.}
\end{figure}

The spin configuration with $c=2$ is shown in FIG.~\ref{fig:sc}, where we make $Q_1=-x, Q_2=y$ and $Q_3=z$. The conventional unit cell with $-2\pi<x,y,z<2\pi$ is chosen in the plot. Cluster of preimages with $n_z=0$ form isolated tori in a simple cubic lattice. Two arbitrary preimages in the same torus wind around each other once after completing a full circle. To confirm the hopfion nature of each texture, we calculate the Hopf index in a way without loss of generality. We calculated for each primitive unit cell as $-\pi<Q_1,Q_2,Q_3<\pi$. Depending on the value of $c$, it is given by
\begin{align}
  Q_{H}=
  \begin{cases}
    \ \ \ 0,\quad |c|>3 \\
    -1,\quad 1<|c|<3.   \\
    \ \ \ 2, \quad 0\le|c|<1
  \end{cases}
\end{align}
Flipping any $Q_i$ odd number of times will change the sign of $Q_H$. The purpose of choosing the parameters as $Q_1=-x,\ Q_2=y$ and $Q_3=z$ is to get a $Q_H=1$ simple cubic hopfion crystal structure in FIG.~\ref{fig:sc}. The large $|c|$ limit is the trivial spin polarized state, so the Hopf index is zero. The details of the calculation are in Supplementary Materials Section I~\cite{supp}. Interestingly, when $0<|c|<1$, each unit cell contains two hopfions with different sizes, as shown in FIG.~\ref{fig:diff-c} where we make $c=\pm{0.5},0$ and $Q_1=x,\ Q_2=y$ and $Q_3=z$. The size of hopfions is alternating layer by layer. The topological index for each of them is $+1$. As $c\rightarrow0$, the two hopfions in each unit cell become equal in size. The lattice is converted to a closely packed body-centered cubic instead.

\begin{figure*}
  \includegraphics[width = \textwidth]{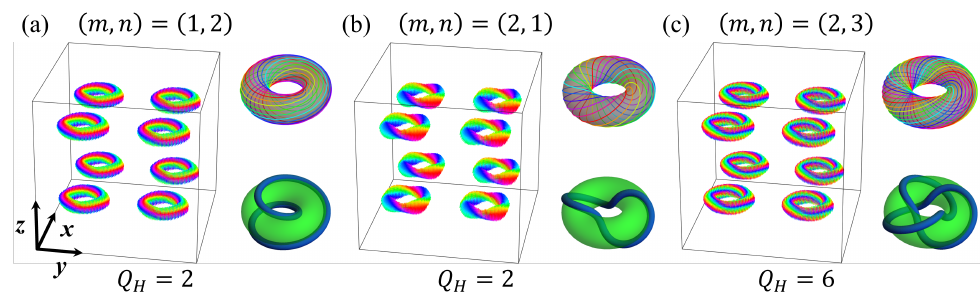}
  \caption{\label{fig:mn}The simple cubic structures of $(m,n)$-tori. (a) $(m,n)=(1,2)$. (b) $(m,n)=(2,1)$. (c) $(m,n)=(3,2)$. The left figure of each one is the simple cubic structure of $(m,n)-$torus. The right top figure is the equi-spin contour lines of one torus in the simple cubic structure and the right bottom figure indicates how the spin fields wind in the corresponding structure.}
\end{figure*}

One can try to write the resulting spin configuration into a multi-Q state as the skyrmion crystal $\bn(\br)=\mathbf{p}\sin(\bk\cdot\br)+\mathbf{q}\cos(\bk\cdot\br)$. As shown in the Supplementary Materials Section II~\cite{supp}, 12 wave vectors are needed to re-build the simple cubic hopfion crystal. Among them 6 wave vectors are associated with Bloch helix with both polarization vectors $\mathbf{p}$ and $\mathbf{q}$ perpendicular to the wave vector $\bk$, so that spins rotate in a plane that is perpendicular to the direction of propagation. 4 of them are associated with N\'eel helix with $\bk$ parallel with one of the polarization vector. In this case, spin rotates in a plane containing the propagation direction. Other than these, there is one component $-2c\hat{z}\sin z$, whose polarization is collinear with the wave vector. This resembles a spin density wave state. This component is responsible for the equivalency between two hopfions in the simple cubic lattice mentioned above. Due to the complexity of this multi-Q decomposition, realization of a hopfion crystal in real materials poses a challenge.  Nevertheless, the analysis of each component above sheds light to the materials design and provides a hint for the experimental detection of hopfion crystals with methods such as neutron and x-ray scattering\cite{muhlbauer_skyrmion_2009,zhang_cdrexs_2018}. Moreover, if any of two $Q_{i}$ are linear dependent, no hopfion is formed. The configurations with different choice of $\phi_{i}$ can be connected by the spatial translation. Some configurations with different combinations of $\phi_{i}(i=1,2,3)$ are shown in Supplementary Materials Section III~\cite{supp}.

Choosing other sets of $\bk_i$-vectors, hopfion crystals with other geometries can also be constructed. One may use $\bu(\bk_i)=\frac{1}{2}(\hat{l}-i\hat{\bk}_i)$. Alternatively, one can continue to use Eq.~\ref{eq:u_vector}. Two approaches just differ by a spin rotation from each other, but the structure of hopfions and the value of the Hopf index are unchanged. We will use the latter approach in what follows.
Selecting ${\bf k}_{1}=\frac{1}{\sqrt{3}}(1,1,1),\ {\bf k}_{2}=\frac{1}{\sqrt{3}}(1,1,-1)$ and ${\bf k}_{3}=\frac{1}{\sqrt{3}}(1,-1,-1)$, the unit vectors of a body-centered cubic lattice, one can yield a face-centered cubic hopfion crystal.
Similarly, choosing ${\bf k}_{1}=\frac{1}{\sqrt{2}}(1,1,0),\ {\bf k}_{2}=\frac{1}{\sqrt{2}}(1,0,1)$ and ${\bf k}_{3}=\frac{1}{\sqrt{2}}(0,1,1)$ yields a body-centered cubic hopfion crystal. These structures are shown in FIG.~\ref{fig:fcc+bcc}.

So far only crystal of hopfions with $Q_H=1$ is concerned. Using complex rational map, our method can be used to generate crystals of other hopfion textures with higher Hopf index, such as axially symmetric tori, torus links, and torus knots.

\begin{figure}[!htbp]
  \includegraphics[width = \columnwidth]{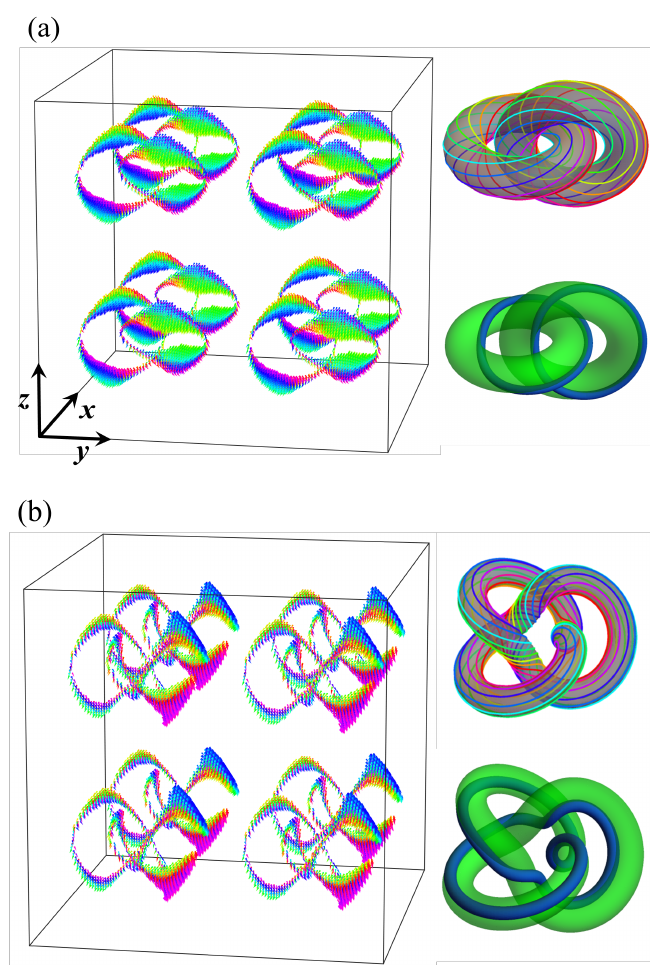}
  \caption{\label{fig:links+trefoil}(a) The simple cubic structure of $Q_H=6$ torus links. (b) The simple cubic structure of $Q_H=7$ trefoils. The left figures of both (a) and (b) are the simple cubic structures. The right top figures are the equi-spin contour lines of one unit cell and the right bottom figures indicate how the spin fields wind in the corresponding structure. }
\end{figure}

The axially symmetric torus is named as $(m,n)$-torus or $\mathcal{A}_{m,n}$, where $m$ and $n$ are the winding numbers of the spin fields in two orthogonal directions. To construct a $\mathcal{\mathcal{A}}_{m,n}$ spin texture, one can choose the Riemann coordinate
\begin{align}
  \xi=\frac{Z_{1}^{m}}{Z_{0}^{n}}.
\end{align}
The resulting structure is a crystal of $Q_{H}=mn$ hopfions. The aforementioned $Q_H=1$ hopfions correspond to $m=n=1$. Several simple cubic hopfion crystals with different combinations of $(m,n)$ are shown in FIG.~\ref{fig:mn}. For simplicity, only the simple cubic form is demonstrated, other forms of the crystal can be constructed in a similar way as discussed above.

For torus links $\mathcal{L}_{n,n}^{1,1}$(the number of subscripts denote the number of linking components, each subscript describes the charge of the linking component, and the superscript denotes the times of linking between different components), the Riemann coordinate is given by
\begin{align}
  \xi=\frac{Z_{1}^{n+1}}{Z_{1}^{2}-Z_{0}^{2}}.
\end{align}
When $n=2$, this results a torus link with $Q_{H}=2n+2=6$, which has two charged 1 torus-like hopfions linked with each. The $Q_{H}=6$ simple cubic crystal is shown in FIG.~\ref{fig:links+trefoil}(a). For even more complex link topologies with additional linking components (e.g., $\mathcal{L}_{n,n,n}^{2,2,2}$), similar rational map formulations can be developed by extending the polynomial expressions in the numerator and denominator.

Finally, the Riemann coordinate of torus knot $\mathcal{K}_{a,b}$ is expressed as
\begin{align}
  \xi=\frac{Z_{1}^{\alpha}Z_{0}^{\beta}}{Z_{1}^{a}+Z_{0}^{b}}.
\end{align}
The trefoil knot is characterized by the parameters as $a=3,b=2,\alpha=2$ and $\beta=1$, yielding a Hopf index $Q_{H}=a\beta+\alpha b=7.$The simple cubic trefoil crystal is illustrated in FIG.~\ref{fig:links+trefoil}(b). Similarly, knot structures with higher Hopf indices can be constructed by modifying the values of $a$, $b$, $\alpha$, and $\beta$. The details of calculating the spin fields through the Riemann coordinate are in Supplementary Materials Section IV~\cite{supp}.

In summary, we have demonstrated a systematic approach for constructing hopfion crystals through the superposition of $\mathbb{R}^4$ helical waves and rational maps. Our framework enables the creation of various cubic lattices (simple cubic, face-centered cubic, and body-centered cubic) with tunable topology. By controlling winding numbers and geometric parameters, we can systematically generate hopfion crystals with different Hopf indices and complex topological structures, including axially symmetric tori, torus links, and torus knots. Our theoretical framework provides a foundation for studying hopfion crystals using variational methods and numerical simulations, which could lead to the identification of real materials that can host hopfion crystals as ground states. For example, it would be interesting to check the energy of hopfion crystals in the Skyrme-Faddeev model. Furthermore, our approach opens avenues for investigating collective behaviors of hopfion ensembles, such as phase transitions, internal modes, magnon bands, and universality classes, which would deepen the understanding of 3D topological solitons and advance the development of hopfion-based devices

W.T Hou and Z. Xiang thank Prof. Yijie Shen for the delighted discussion on structures of hopfions. We are grateful for fruitful discussions with Prof. Naoto Nagaosa. 
\bibliography{short, Hopfion_Crystal}

\end{document}


\title{Construction of Hopfion Crystal\\
--- Supplementary material ---}

\author{Wen-Tao Hou}
\author{Zhuoxian Xiang}
\affiliation{School of Physical Science and technology, Tiangong University, Tianjin 300387, China}

\author{Yizhou Liu}
\affiliation{Anhui Province Key Laboratory of Low-Energy Quantum Materials and Devices, High Magnetic Field Laboratory, HFIPS, Chinese Academy of Sciences, Hefei 230031, China}

\author{Jiadong Zang}
\affiliation{Department of Physics and Astronomy, University of New Hampshire, NH 03824, USA}

\date{\today}

\maketitle

\section{Calculation of Hopf index }

For a general case, we define the spinor field without normalization
as $|\chi_{0}\rangle=\frac{1}{\mathcal{N}}|\chi\rangle$ where $\mathcal{N}$
is the normalization factor in Eqn.(12) from the main text      . So we have
\begin{align}
{\bf a} & =-i\langle\chi|\nabla|\chi\rangle\nonumber \\
 & =-i\langle\chi_{0}|\mathcal{N}\nabla(\mathcal{N}|\chi_{0}\rangle)\nonumber \\
 & =-i\mathcal{N}^{2}\langle\chi_{0}|\nabla|\chi_{0}\rangle-i\mathcal{N}\nabla\mathcal{N}\langle\chi_{0}|\chi_{0}\rangle\nonumber \\
 & =\mathcal{N}^{2}{\bf a}_{0}-i\frac{\mathcal{\mathcal{N}\nabla}\mathcal{N}}{\mathcal{N}^{2}}\nonumber \\
 & =\mathcal{N}^{2}{\bf a}_{0}-i\nabla\ln\mathcal{N}
\end{align}
where ${\bf a}_{0}=-i\langle\chi_{0}|\nabla|\chi_{0}\rangle$. By
a $U(1)$ gauge transformation ${\bf a}\rightarrow{\bf a}+i\nabla\ln\mathcal{N}$,
${\bf a}$ is equivalent to $\mathcal{N}^{2}{\bf a}_{0}.$ The gauge transformation works because $\mathcal{N}$ is well defined. Then we
have 
\begin{align}
{\bf b} & =\nabla\times{\bf a}\nonumber \\
 & =\nabla\times(\mathcal{N}^{2}{\bf a}_{0})\nonumber \\
 & =\mathcal{N}^{2}\nabla\times{\bf a}_{0}+\nabla\mathcal{N}^{2}\times{\bf a}_{0}.
\end{align}
So the density of Hopf index is 
\begin{align}
h & ={\bf b}\cdot{\bf a}\nonumber \\
 & =\mathcal{N}^{4}{\bf a}_{0}\cdot(\nabla\times{\bf a}_{0})+\mathcal{N}^{2}{\bf a}_{0}\cdot(\nabla\mathcal{N}^{2}\times{\bf a}_{0})\nonumber \\
 & =\mathcal{N}^{4}{\bf a}_{0}\cdot{\bf b}_{0}
\end{align}
in which ${\bf b}_{0}=\nabla\times{\bf a}_{0}$. Then the Hopf index
of one unit cell in the hopfion crystal can be calculated simply by
\begin{align}
H & =-\frac{1}{4\pi^{2}}\int_{V}hdV\nonumber \\
 & =-\frac{1}{4\pi^{2}}\int_{V}\mathcal{N}^{4}{\bf a}_{0}\cdot {\bf b}_{0}dV.
\end{align}
where $V$ is the volume occupied by one unit cell in the hopfion
crystal.

\section{$\phi_{i}$ dependence}

To illustrate the $\phi_i$ depedence, we make the parameters of Eq.12 from the main text as $Q_{1}=x+\phi_{1},\ Q_{2}=y+\phi_{2}$,\ $Q_{3}=z+\phi_{3}$ and $c=2$. The choice will result in  simple cubic hopfion crystal structures. Then we plot the hopfion crystal structures with several different combinations of $\phi_{i}(i=1,2,3).$ They are shown in FIG.1. It is obvious that configurations with different combinations can be connected by spatial transformations.

\begin{figure}[!htbp]
\includegraphics[scale=0.75]{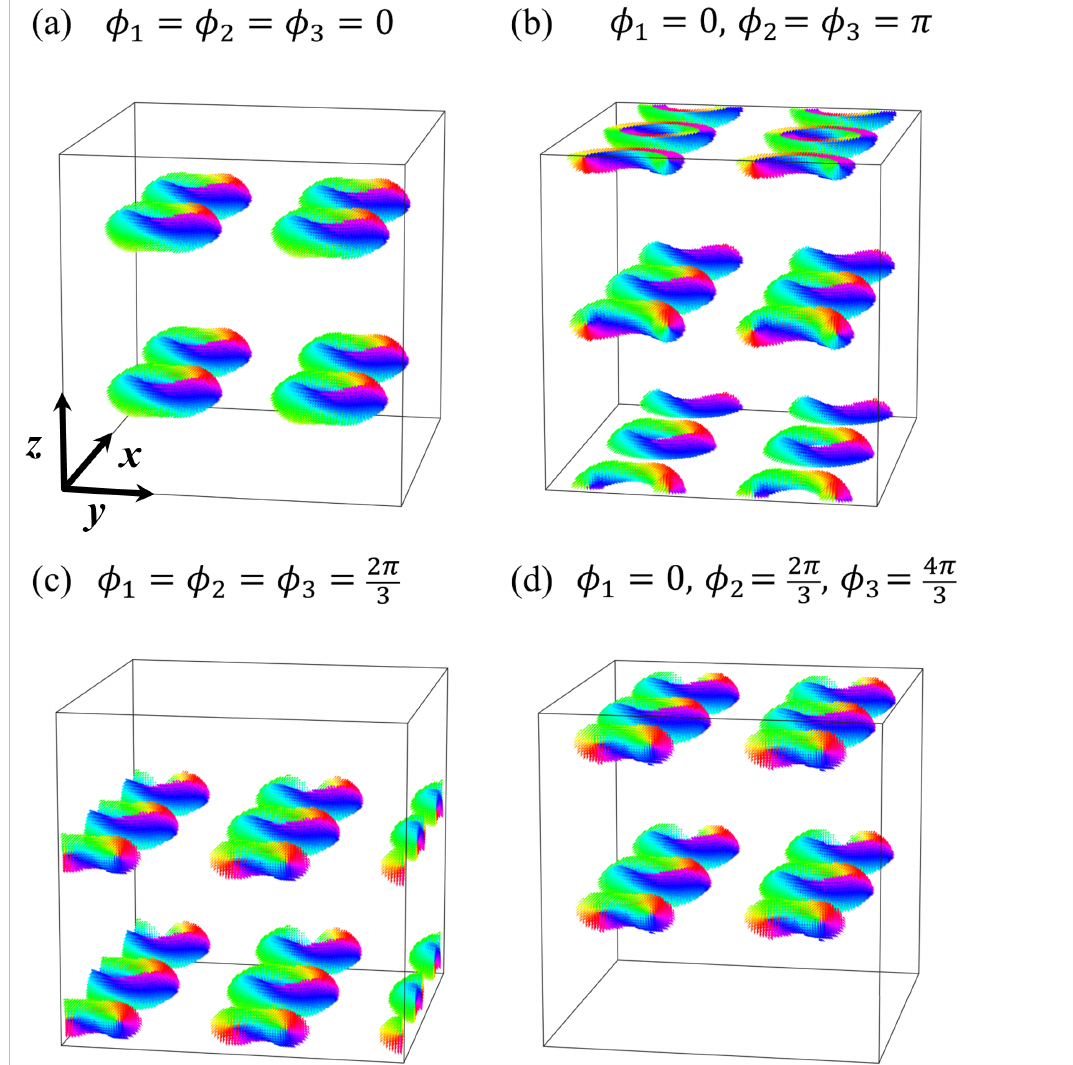}
\caption{The different combinations of $\phi_{i}(i=1,2,3).$ We make $c=2$ and $x,y,z$ are from $-2\pi$ to $2\pi$. }
\end{figure}

\section{The components of Hopfion crystal structures}

${\bf {n}}$ is the spin texture without normalization which is given by Eq.12 from the main text. It can be decomposed into the form 
\begin{equation}
\mathbf{n}(\mathbf{r})=\sum_{m,n,k}{\bf p}_{mnk}\cos(mQ_{1}+nQ_{2}+kQ_{3})+{\bf q}_{mnk}\sin(mQ_{1}+nQ_{2}+kQ_{3}).
\end{equation}
The details are shown in TABLE I. Totally, there are 12 wave vectors. For simplicity, we make $Q_1=x,Q_2=y$ and $Q_3=z$. Among them, 6 wave vectors are associated with Bloch helix with both polarization vectors $\mathbf{p}$ and $\mathbf{q}$ perpendicular to the wave vector $\bf k$, so that spins rotate in a plane that is perpendicular to the direction of propagation. 4 of them are associated with N\'eel helix with $\bf k$ parallel with one of the polarization vector. In this case, spin rotates in a plane containing the propagation direction. Other than these, there is one component $-2c\hat{z}\sin z$, whose polarization is collinear with the wave vector. When $c=0$, the spin density wave component, which has the helix ${\bf p} _{001}=-2c\hat{z}$ and ${\bf q} _{001}=0$, and $m=0,n=1,k=0$ helix, which has ${\bf p} _{001}=-2c\hat{z}$ and ${\bf q} _{001}=2c\hat{x}$, will disappear. Accordingly, the periodicity of the spin texture turns out to be half of the ones when $c\neq0$. The configurations of simple cubis structures with $c=\pm0.5$ and $c=0$ are shown in FIG. 3 from the main text. Obviously, when $c=\pm 0.5$, the hopfions in these crystal structures have two sizes. And when $c=0$, the hopfions have only one size. 

\begin{table}[!htp]
\begin{centering}
\begin{tabular}{|c|c|c|c|c|c|}
\hline 
$m$  & $n$  & $k$  & ${\bf p}_{mnk}$  & ${\bf q}_{mnk}$  & spin texture\tabularnewline
\hline 
\hline 
2  & 0  & 0  & $-\hat{z}$  & $\hat{y}$  & Bloch helix\tabularnewline
\hline 
1  & 1  & 0  & $-\hat{z}$  & $\hat{x}+\hat{y}$  & Néel helix\tabularnewline
\hline 
1  & 0  & 1  & $-\hat{x}-\hat{z}$  & $\hat{y}$  & Néel helix\tabularnewline
\hline 
1  & 0  & 0  & $-2c\hat{z}$  & $2c\hat{y}$  & Bloch helix\tabularnewline
\hline 
1  & 0  & -1  & $\hat{x}-\hat{z}$  & $\hat{y}$  & Néel helix\tabularnewline
\hline 
1  & -1  & 0  & $-\hat{z}$  & $-\hat{x}+\hat{y}$  & Bloch helix\tabularnewline
\hline 
0  & 2  & 0  & $-\hat{z}$  & $\hat{x}$  & Bloch helix\tabularnewline
\hline 
0  & 1  & 1  & $\hat{y}-\hat{z}$  & $\hat{x}$  & Bloch helix\tabularnewline
\hline 
0  & 1  & 0  & $-2c\hat{z}$  & $2c\hat{x}$  & Bloch helix\tabularnewline
\hline 
0  & 1  & -1  & $-\hat{y}-\hat{z}$  & $\hat{x}$  & Néel helix\tabularnewline
\hline 
0  & 0  & 1  & $-2c\hat{z}$  & 0  & spin density wave\tabularnewline
\hline 
0  & 0  & 0  & $(-1-c^{2})\hat{z}$  & 0  & polarized\tabularnewline
\hline 
\end{tabular}
\par\end{centering}
\caption{The components of spin texture ${\bf S}$.}
\end{table}

\section{Rational Map and Riemann Coordinate}

The Riemann coordinate is 
\begin{equation}
W=\frac{n_{1}+in_{2}}{n_{3}+1}
\end{equation}
in which $(n_{1},n_{2},n_{3})$ is the spin vector with the length
is $1$. An intermediate state for it when doing rational map is that
\begin{equation}
W=\frac{a+ib}{c+id}
\end{equation}
in which $a,b,c$ and $d$ are real numbers. So we have 
\begin{equation}
W=\frac{(a+ib)(c-id)}{c^{2}+d^{2}}=\frac{ac+bd+i(bc-ad)}{c^{2}+d^{2}}
\end{equation}
Then we parameterize them as 
\begin{align}
ac+bd & =\rho n_{1},\\
bc-ad & =\rho n_{2},\\
c^{2}+d^{2} & =\rho n_{3}+\rho
\end{align}
where $\rho>0$ . Then we have 
\begin{align}
n_{1} & =\frac{2a(c^{2}+d^{2})}{(c^{2}+d^{2})^{2}+a^{2}+b^{2}},\\
n_{2} & =\frac{2b(c^{2}+d^{2})}{(c^{2}+d^{2})^{2}+a^{2}+b^{2}},\\
n_{3} & =\frac{(c^{2}+d^{2})^{2}-a^{2}-b^{2}}{(c^{2}+d^{2})^{2}+a^{2}+b^{2}}
\end{align}
and 
\begin{equation}
\rho=\frac{(c^{2}+d^{2})^{2}+a^{2}+b^{2}}{2(c^{2}+d^{2})}.
\end{equation}